# ARMrayan Multimedia Mobile CMS: a Simplified Approach towards Content-Oriented Mobile Application Designing

Ali Reza Manashty, Mohammad Reza Ahmadzadeh Raji, Zahra Forootan Jahromi, and Amir Rajabzadeh

***Abstract*—**The ARMrayan Multimedia Mobile CMS (Content Management System) is the first mobile CMS that gives the opportunity to users for creating multimedia J2ME mobile applications with their desired content, design and logo; simply, without any need for writing even a line of code. The low-level programming and compatibility problems of the J2ME, along with UI designing difficulties, makes it hard for most people –even programmers- to broadcast their content to the widespread mobile phones used by nearly all people. This system provides user-friendly, PC-based tools for creating a tree index of pages and inserting multiple multimedia contents (e.g. sound, video and picture) in each page for creating a J2ME mobile application. The output is a stand-alone Java mobile application that has a user interface, shows texts and pictures and plays music and videos regardless of the type of devices used as long as the devices support the J2ME platform. Bitmap fonts have also been used thus Middle Eastern languages can be easily supported on all mobile phone devices. We omitted programming concepts for users in order to simplify multimedia content-oriented mobile applictaion designing for use in educational, cultural or marketing centers. Ordinary operators can now create a variety of multimedia mobile applications such as tutorials, catalogues, books, and guides in minutes rather than months. Simplicity and power has been the goal of this CMS. In this paper, we present the software engineered-designed concepts of the ARMrayan MCMS along with the implementation challenges faces and solutions adapted.

## I. Introduction

THE trend towards the wide use of mobile phones as a must-have company, draws the attention of manufacturers and developers to enhance the capabilities of these devices and bring more functions and facilities from PCs and other regular devices to mobile phones.

Most of the current cell phones are also a good digital camera, a MP3/MP4 player, a web browser and a game console. Although various types of application have been written for these devices, there's still a huge difference between the capabilities of the PC applications and the Mobile ones. This difference is caused by the lack of processing capabilities and the libraries written for these devices.

There are various platforms and mobile operating systems currently available for mobile phones, such as Java, Symbian OS and Windows Mobile. Java supported platforms are becoming more common, as in recent Symbian OS based handsets, J2ME applications are also supported as a side platform. One of the most promising software platforms for mobile devices is Java 2 Micro Edition. Sun representatives asserted that 18 to 20 million mobile phones support the J2ME platform (in the year 2003) and analysts prediction that within the next few years, this technology will become omnipresent, and Gartner Group estimates that, in 2006, approximately 80 percent of mobile phones will support Java [1], has become a reality. The low-level programming and compatibility problems of the J2ME, along with UI (user interface) designing difficulties, make it hard for most programmers to develop applications for this platform. One distinguishable attempt to enhance the UI and the utility libraries of the J2ME platform for the *programmers* is J2ME Polish [2]. Despite major utilities a programmer can use to make his/her job easier in J2ME programming by using J2ME Polish, there are still difficulties using this tool (e.g., IDE integration). Although a book has been published about J2ME Polish [3], but there are still problems using it that makes it double hard for developers to adopt either base J2ME platform libraries or the libraries and UI presented in J2ME Polish.

Nowadays people spend a notable amount of time in transportation, without having access to their PCs or having hard time accessing their labtops; instead, they are constantly using their cell-phones/PDAs. Because of this, many companies and active advertisement centers, whether religious, educational or business related are pinpointing on these handset devices for applying their policies. Despite of this huge trend, there is still a lack of programmers that write both

Dr. Amir Rajabzadeh is with Razi University, Computer Engineering Faculty, Bagh-e-Abrisham, Iran, Kermanshah (Mobile: +989123473189; Tel/Fax: +988314283265; e-mail: rajabzadeh@razi.ac.ir).
Ali Reza Manashty is with Razi University, Computer Engineering Faculty, Bagh-e-Abrisham, Iran, Kermanshah (Corresponding author) (Mobile:+989355332577; fax: +988318359105; e-mail: a.r.manashty@gmail.com)
Zahra Forootan Jahromi is with Razi University, Computer Engineering Faculty, Bagh-e-Abrisham, Iran, Kermanshah (Mobile: +989177922901; e-mail: zahra.forootan@gmail.com)
Mohammad Reza Ahmadzadeh Raji is with Razi University, Computer Engineering Faculty, Iran, Kermanshah e-mail: moh853@gmail.com)

visually and functionally acceptable mobile applications in beneficial time and cost. Some people worked on the importance of the GUI [4] while others focused on the commercial and educational importance of mobile applications [5], [6], [7], [8]. Still there is no common tool for users to create their desired mobile-based content that can be used by a wide variety of people during their daily life.

In web-based services, the term CMS (Content Management System) is used to describe a web application that gives content-oriented accessability to the administrator of the website, regardless of his/her knowledge in programming concepts of the website. In addition to content-oriented access, the administrator can also change the visual theme of his/her website. Joomla [9] and Wordpress [10] CMSs are open-source samples of so called website applications. In PC applications scope, there are lots of programs that help users create multimedia applications without any need of software designing challenges. Now that mobile devices are undetachable part of everybody's routine; there is still a need for mobile content creation tools so that the providers of contents can publish their data (whether rich-content data or simple texts) through structured theme-based applications that can also provide some facilities for using mobile device special capabilities (e.g., sending contents through SMS/MMS or media playback; that using mobile browsers for these jobs is still facing some difficuties). In this paper, we'll present a solution for a simplified approach toward content-oriented mobile application designing.

## II. ARMrayan Multimedia Mobile CMS

### A. Preface

As mentioned in the previous section, there's still a need for a system that provides tools for easy management of contents for use in mobile application providing additional mobile device facilities for efficient distributing or viewing of mobile-based contents. Such applications may now be called Mobile Content Management Systems or MCMSs and talk about the ARMrayan Multimedia MCMS design and implementation concepts in the following sections.

### B. Overview

The ARMrayan Multimedia MCMS is a PC-based application providing the user the tools needed for inserting his/her desired multimedia content in a tree-based index of pages, in addition to providing simple theme-based UI designing facilities. The output of this Windows application is a J2ME mobile application with the desired content and visual design. The J2ME application was developed separately using the Java language for MIDP2.0+ supporting mobile phones. This base J2ME file that the Windows program recieves as input is called the Jar Template File. This Template File can be modified or updated regularly to maintain compatibility with newer devices and/or get improvement in the UI or search engine features.

After the user designed the tree index of pages in the Windows program, he/she will now be able to add several contents per page in the desired order. These content types can currently be among texts, pictures, sounds or music, videos and animations, map points, phone number, e-mail address, web sites link and etc.

### C. Design

The design process of this software uses an evolutionary model in which the design and implementing the software are joined together while developing early versions of the software. The reason is that although some original ideas are considered as the basis of the project objectives, the implementation can affect the milestones during the development of the software. Many original ideas for developing the software will be out of order as the desired tools or libraries expected will not be available. This problem affects wide-range software development using J2ME as the final methods and approaches initially estimated fails regularly as more limitations of the MIDP appears during development process.

The UML diagrams are designed for different part of the application where the final mobile application user (end-user) and the Windows application user are interacting with the system. The use-case diagram of the possible ways the final mobile application user can act with the mobile software is illustrated in Fig. 1.

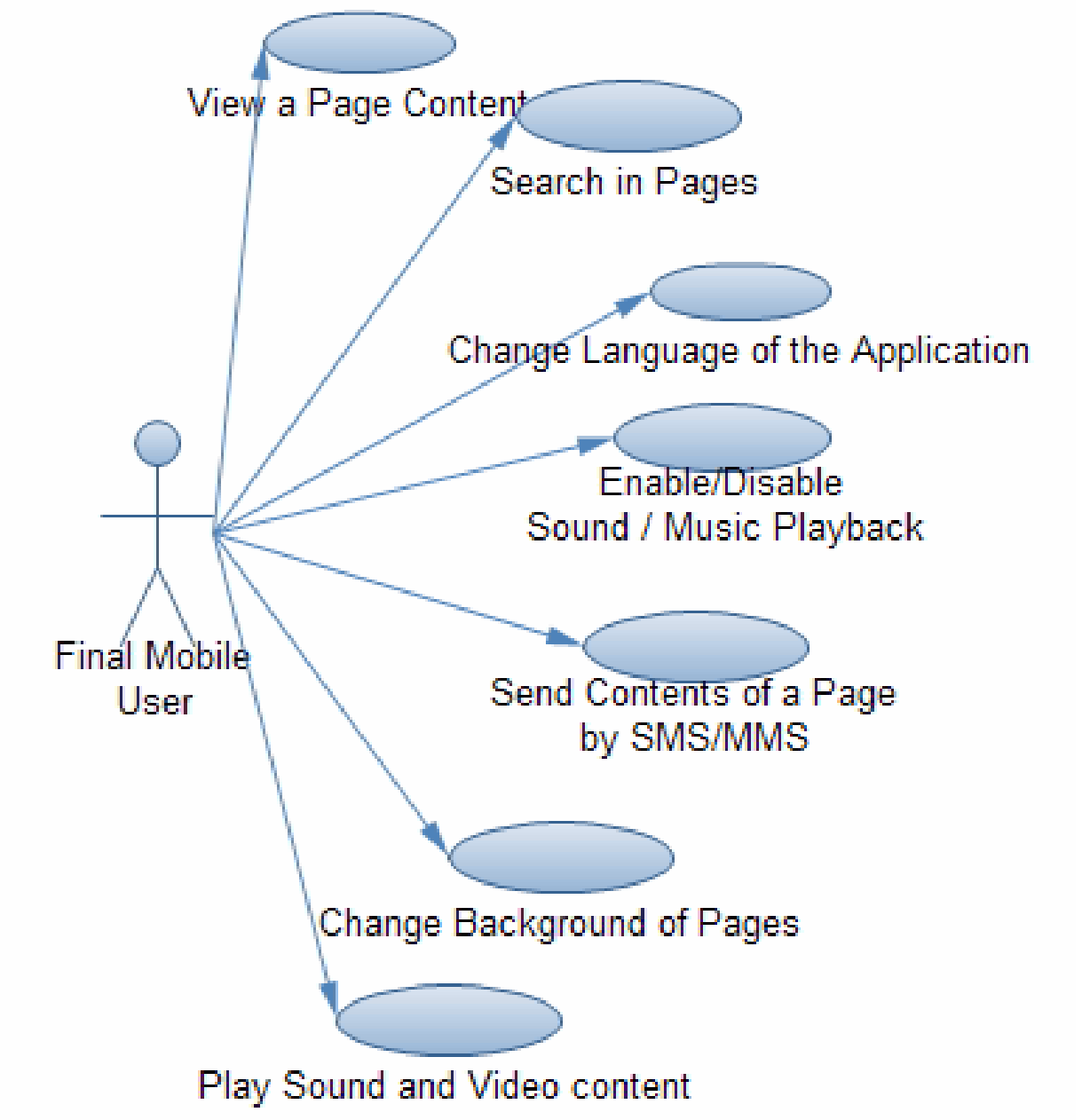

Fig. 1 The use-case diagram of the final mobile application exported by the ARMrayan Mobile CMS

One the most important non-functional requirement of the main application is considered the user friendliness of the user interface that manages retrieving the data from the user and generating the contents. The ease of use is another non-functional requirement that was focused during design and implementation. It was created in two different languages and supports hot swapping of the languages of the whole program while running it.

The visual features of the final mobile application, like colors and backgrounds, which were chosen by the user, are defined in binary setting files for the mobile engine to read from. This novel approach to mobile content management systems makes it extremely easy for the public to create their own mobile books, catalogues and similar mobile applications.

Apart from text, other media can easily be inserted in each page. Sounds are automatically detected in the output program. If the end user comes across a page that should contain sounds, the sound is automatically detected and played if it's representing icon is in the screen's visibility range. If two sounds are placed in one page, the program automatically selects the first and can toggle to the next sound with just a click. Videos and animations also work the same and the user can choose the one he/she wants to play. With this method it will be easy to create mobile multimedia tutorials and books.

The mobile engine is designed to be selected in the main application as an external file, thus making the mobile software easily updatable. With the development of new features in the mobile core, visual features can be added and different templates can be created over time. This flexibility of the program makes easier development of the project in the future. In Fig. 2 the use-case diagram of the main program is also illustrated.

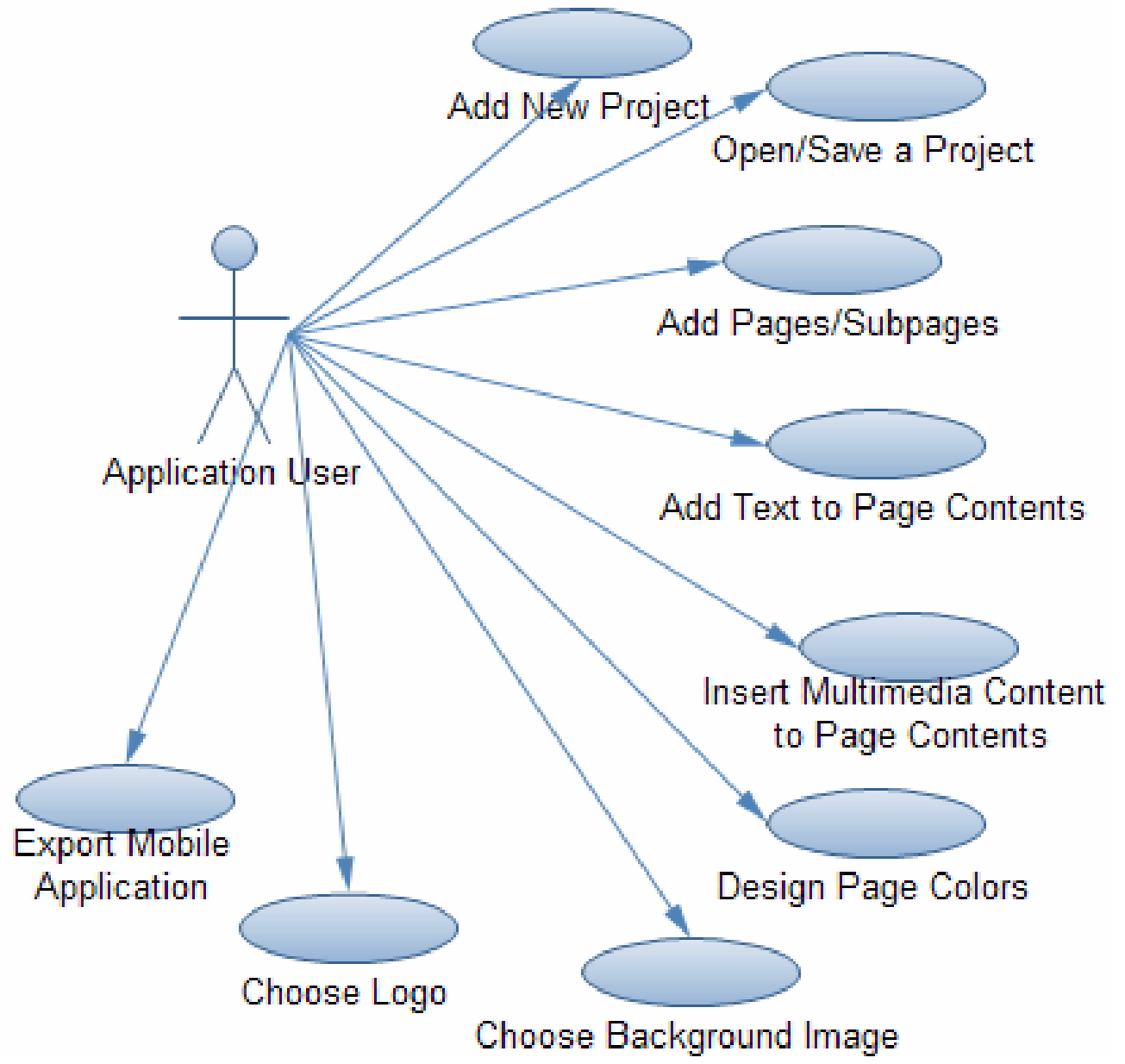

Fig. 2 The use-case diagram of the main engine application-ARMrayan Mobile CMS

### D. *Implementation*

Implementation of the Windows application was using Visual C# .Net while the mobile engine application was developed in Java 2 Micro Edition language. The main features of the implemented software are as following:

1. Supporting texts, sounds, images, videos, phone numbers, web links and emails as the input content.
2. Tree-view page designing.
3. Displaying Middle Eastern fonts via the bitmap font framework.
4. Multilanguage environment (English and Farsi).
5. Environment theme designing.
6. Sending contents via SMS/MMS.
7. Text searching.

Detailed information regarding the implementation of the ARMrayan MCMS is presented in the following sections.

#### *1) The implementation of the main engine*

The program itself is simply the interface between the user and the core mobile engine. The program retrieved the content from the user and generates the output J2ME application. This is done by extracting the mobile engine file, inserting the binary content files within the template, and recompressing the Jar file and giving it to the user as the output. It is good to mention that the program runs the output file with an emulator to show it to the user as a preview ergo making it even more user-friendly.

The main program engine displays the content in tree views thus making it possible for a page to have innumerous sub pages (Fig. 3). Each node of the tree index is a page containing the contents. It is also possible to change the order of the contents within each page. The colors, background pictures and the background music of different areas of the output program can also be set in the main user interface. Again a preview is possible for such settings.

#### *2) The implementation of the mobile engine*

The mobile engine of the ARMrayan MCMS was developed as J2ME application. This engine is actually a stand-alone mobile application simply known as a "Template File" in the application, containing the main algorithms for displaying or playing different contents such as sounds, videos and texts.

This mobile engine was designed with the use of the Canvas class of the J2ME platform. The Canvas was used for many reasons. Amongst those of importance are the low level privileges the Canvas offers for visual design. Also the use of bitmap fonts is only possible with the use of Canvas class.

This mobile software (Fig. 4) also contains the functions to show the menus and performing other operations such as searching within the content and handling the bitmap font for Asian languages. The contents are fed to the template file via binary files generated by the main program. The mobile engine's main job is to read the data from the binary files and to display the read data in the appropriate pages. The first binary file that this core processes is the file containing the index of the contents. One use of the index file is to show the index of contents to the user and the second use, being quite invisible to the user, is the fact that the index file contains the byte position of each page within the main content file thus making it possible to retrieve the page without the need of reading the whole file. This feature which uses the SkipByte() method of the J2ME platform makes it faster to read large content files (J2ME file system does not support seeking). Mobile phones processing and I/O capabilities are limited, therefore skipping the unwanted bytes before the selected page's content is considered as a very significant improvement.

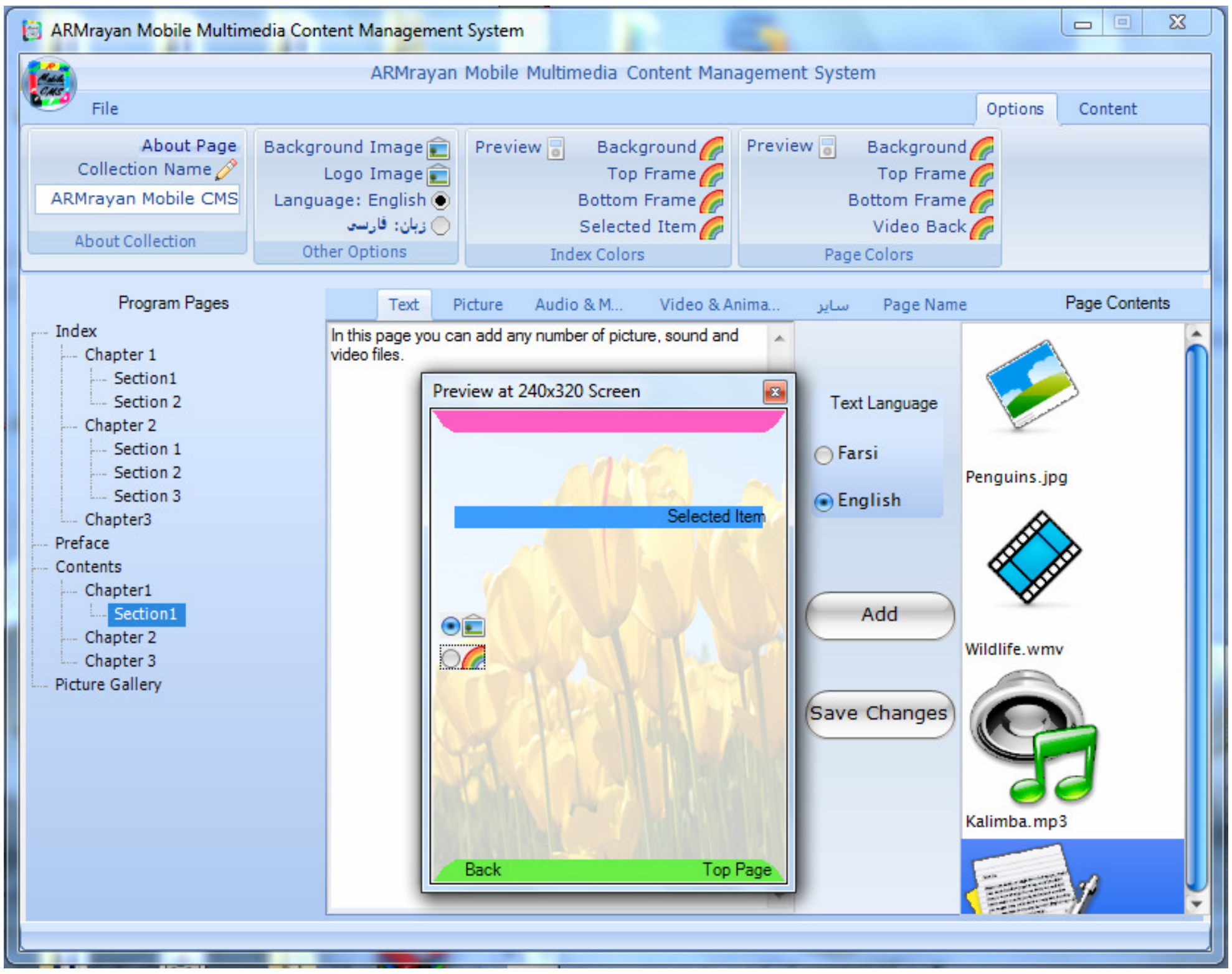

Fig. 3 The ARMrayan Mobile Multimedia CMS core engine. This image shows a list of pages on the left, the content of the selected page on the right and the preview of the designed application in the center of the screen.

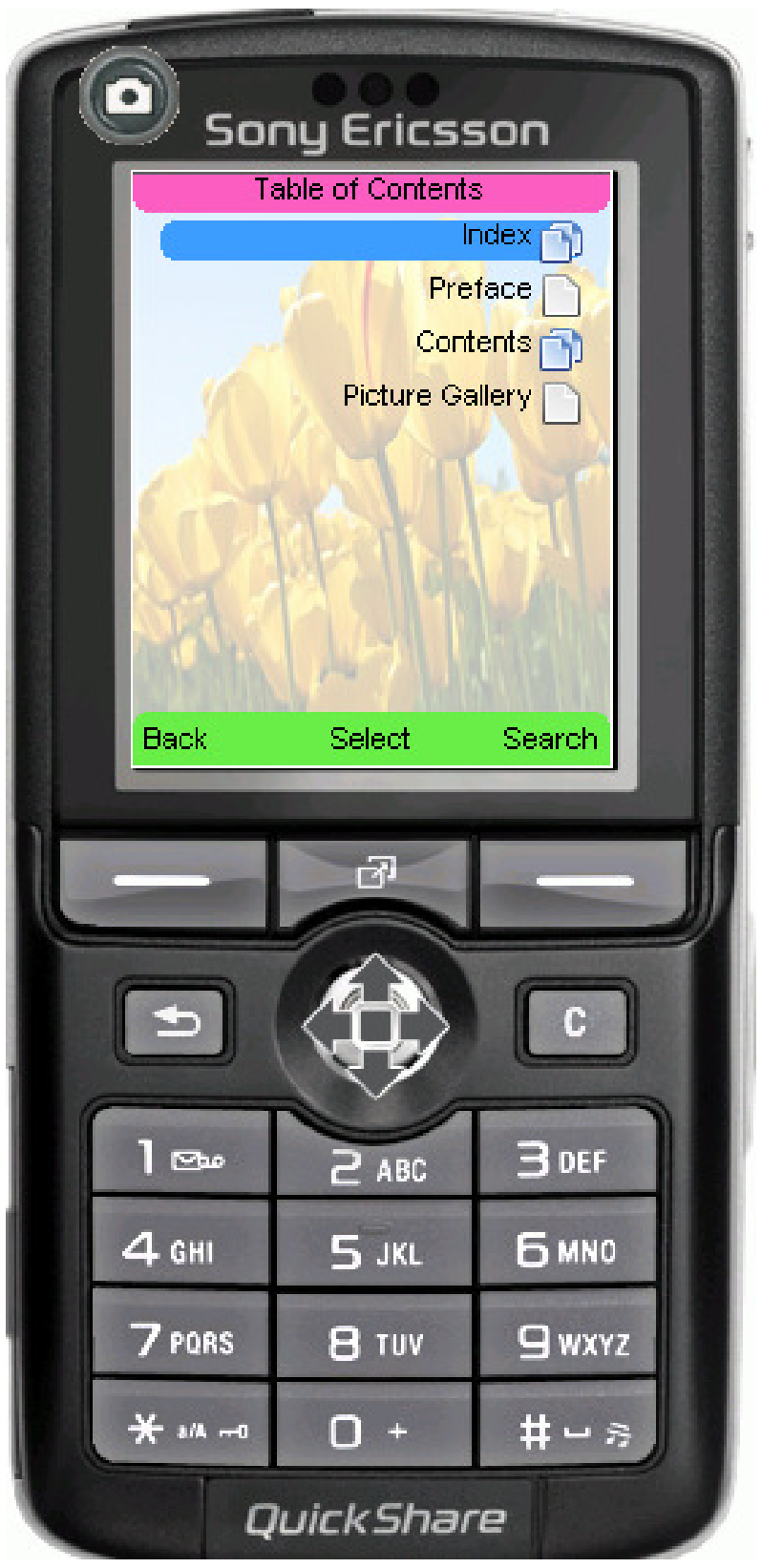

Fig. 4 The final mobile application view. This page is the output J2ME application of the project in illustrated in Fig. 3.

The font is also retrieved by the mobile engine as a bitmap image. The mobile engine file retrieves the text contents, finds the Asian letters used in the bitmap image and displays them. This makes displaying Asian fonts on non-Asian supported mobile phones possible. With this implementation of non-English fonts, multiple fonts with different colors can be used within the content.

This approach for using unsupported fonts is done by extracting alphabet of the language to an image. Rebuilding the word and sentences as images using the alphabet generated earlier in the font image file, solves this problem. The reason why J2ME platforms does not support regular font types (e.g., TTF) is the lack of processor speed and libraries needed for processing regular curve data of the True Type Fonts.

Some other configurations are also read by the mobile engine, such as whether splash pages should appear or not, or the address of contents other than texts like sound files or videos. Sounds and video files are played using the methods defined in the MIDP2.0 platform.

## I. Challenges Faced

Maybe the most challenging of all tasks is capability of writing Asian languages in mobile phones that do not support them. Although bitmap fonts have been used before in other languages such as English but there have been little approaches in Middle Eastern languages that can be considered complete or flawless. The reason is because of the complex way that such languages are written. There are cases when a letter's form and shape depends on the letters before and after it. In some cases letters must stick together and sometimes not. Our approach was considerably well formed. The alphabet of the language, in this case the Farsi and Arabic alphabet, was divided into two groups. The first contains letters that can stick to other letters and the second group contains letters that do not stick at all. Symbols and numbers are considered in the latter group. While writing the text from the binary files the program checks whether a letter can stick at all or not, if it can then it checks the letters before and after it to find the exact shape of the letter and then it draws the letter in the canvas. The size of the letters drawn until then is checked before each draw to make sure the line doesn't go on more than the width of the mobile screen. If it does then the sentence is continued in the second line.

## II. Conclusion

In this paper we presented a content management system for mobile devices that support the J2ME platform. The unique idea of the ARMrayan MCMS is its content-oriented structure which grants the user full control over the content he/she may want to share and still without the need of writing even a line of mid level J2ME language code. The design of the mobile engine as a core and the insertion of the user's data within a template file is the main key of this solution. Features like the ability to write in Middle Eastern languages like Farsi and Arabic and the ability to change the fonts are amongst the unique abilities that make this content management system novel. Above all the features of the software, the idea to bring together the popular multimedia contents in the mobile phone applications is of the highest importance, something that we believe is necessary in the current decade.

In the first month of developing this system, we awarded the 1st Place in the 3rd International Digital Media Festival as the "The Best Mobile Software in Technology, Innovation and Development" for "ARMrayan Mobile CMS"; Iran-Tehran; October 2009 [11]. Not many days passed, that an output of the program which contained audio and photo gallery, multi-language text and fabulous design, also awarded the 3rd Place in the 4th National Imam-Reza Festival, Professional Mobile Software Title of the Digital Media section, for the "Imam Reza Pilgrim Mobile Software"; Iran-Tehran; December 2009 [12]. We also sponsored by a high cultural and religion center for upgrading the application with their desired need, so that they can produce multi-language cultural and religious content-oriented mobile application for broadcasting into the society.

## III. Future Works

It is well known that improving and updating such projects or ideas is highly important, and making the program to be easily flexible towards improvement and future development, is considered of even higher importance. Fortunately this software is completely designed and implemented for future extensions. The mobile engine can be upgraded and improved regardless of modifying the main program. The main program itself remains a powerful, yet simple base method for inserting contents in different mobile engine.

Other contents that are looked forward to be supported in the future are panorama images and map points which will be available for inserting as content. These contents will be inserted in any place within a page and can be selected, captioned and illustrated.

The implementation of maps is one of the main candidates for future development. With the implementation of this ability users will easily be able to import a map image into the main program as an input, and the program will be able to zoom and move within the map. What makes this feature significant is the ability to add map points anywhere in the page contents captioned and selectable.

In near future works we hope to be able to import panorama images and show them as 360 degrees panoramas with the help of the m3g model file format. Each image is then inserted as the texture of a sphere created from the m3g file and a camera is inserted in the center of the sphere allowing the user to rotate it.

## Acknowledgment

Ali Reza Manashty and Zahra Forootan Jahromi want to thank Mr. Amir Sayah for his great support of us in the 2nd and 3rd International Digital Media Festival. We wouldn't be here, without the help of our families and his perfect company.